\documentclass[aps,onecolumn,amssymb]{revtex4}
\usepackage{epsfig,amssymb,amsmath}

\newcommand{\be}{\begin{equation}}
\newcommand{\ee}{\end{equation}}

\begin{document}
\title{Level statistics for quantum $k$-core percolation}
\author{L. Cao$^1$ and J. M. Schwarz$^1$}
\affiliation{$^1$Physics Department, Syracuse University, Syracuse, NY 13244}
\date{\today}
\begin{abstract}
Quantum $k$-core percolation is the study of quantum transport on
$k$-core percolation clusters where each occupied bond must have
at least $k$ occupied neighboring bonds. As the bond occupation probability, $p$, is increased from zero to unity, the system undergoes a transition from an insulating phase to a metallic phase.  When the lengthscale for the disorder, $l_d$, is much greater than the coherence length, $l_c$, earlier analytical calculations of quantum conduction on the Bethe lattice demonstrate that for $k=3$ the metal-insulator transition (MIT) is discontinuous, suggesting a new universality class of disorder-driven quantum MITs. Here, we numerically compute
the level spacing distribution as a function of bond occupation
probability $p$ and system size on a Bethe-like lattice. The level spacing analysis suggests that for $k=0$,
$p_q$, the quantum percolation critical probability, is greater
than $p_c$, the geometrical percolation critical probability, and
the transition is continuous.  In contrast, for $k=3$, $p_q=p_c$ and the
transition is discontinuous such that these numerical findings are consistent with our
previous work to reiterate a new universality class of disorder-driven quantum MITs.
\end{abstract}

\maketitle

\section{Introduction}

With the exception of transition metal compounds, there exist two conventional paradigms for metal-insulator transitions (MITs): a Mott-Hubbard-type transition and an Anderson-type transition.  The former is a consequence of tuning the interactions between electrons by changing the distance between atoms, for instance, and the MIT is discontinuous~\cite{mott}. The latter is a consequence of the tuning the disorder in the material, and the MIT is continuous~\cite{anderson,mirlin}. Recently, a discontinuous disorder-driven MIT has been predicted for a model in which there exists geometric constraints on the disorder~\cite{liang.jen}. The model has been dubbed quantum $k$-core percolation. Quantum $k=0$-core percolation, or quantum percolation (QP), has been studied since the 1980's and exhibits a continuous, Anderson-type MIT~\cite{degennes,germans2}. Quantum percolation is defined accordingly. Consider a lattice whose bonds are occupied independently and at random with bond occupation probability
$p$. An electron can only hop between lattice sites $i$ and $j$ along an occupied bond and cannot hop along an unoccupied bond.  In addition, there exists a constant on-site binding energy which is set to zero for convenience. In total, the tight-binding Hamiltonian for quantum percolation is \be
H=\sum_{i,j}t_{ij}a_i^{\dagger}a_j+h.c.
\ee\label{eq:tight-binding} in which, \be t_{ij} = \left\{
\begin{array}{ll}
1 & \textrm{with probability $p$}\\
0 & \textrm{with probability $1-p$}
\end{array} \right.\label{eq:tij}
\ee and $a_i^{\dagger}$ and $a_i$ are electron creation and
annihilation operators.

Quantum percolation exhibits a MIT as $p$ is increased from zero,
at least for three dimensions and
above~\cite{germans1,travenec,kaneko}. In two dimensions, some
studies indicate a
transition~\cite{meir,chang,islam1,islam2,germans3}, while others
do not~\cite{grest,mookerjee}. Analytical work on the Bethe
lattice indicates that the quantum percolation transition may be
in the same universality class as geometric percolation except
with the transition probability $p_q$ above which there exists
extended zero-energy wavefunctions is greater than the threshold
above which there exists a spanning cluster,
$p_c$~\cite{harris0,harris}. The geometric percolation transition
is a continuous and, therefore, so is the quantum percolation
transition presumably.

Recently, we have studied quantum $k$-core percolation
on the Bethe lattice~\cite{liang.jen}.  The term $k$-core refers to a geometrical constraint where every occupied bond must have at least $k-1$ occupied
neighboring bonds~\cite{clr,wormald,network0,network,slc,harris.schwarz}. Such a geometric constraint may have implications for glassy systems~\cite{biroli}, jamming systems~\cite{slc}, and even biological systems~\cite{levine}. To enforce this constraint, bonds are initially
occupied independently and at random with probability $p$. Then,
those occupied bonds with less than $k-1$ occupied neighboring
bonds are rendered unoccupied. This removal procedure proceeds
recursively throughout the lattice until all occupied bonds
satisfy the $k$-core constraint. See Fig. 1
for an example on the Bethe lattice.  For $k\ge 3$, the geometric percolation transition is in a different universality from ordinary geometric percolation with, for example, the fraction of occupied bonds in the spanning $k$-core cluster, $P_{\infty}$, scaling with $p$ as $P_{\infty}=P_0+P_1(p-p_c)^{1/2}$ for $p\ge p_c$ (with $p_c-p<<1$) where $P_0$ and $P_1$ are constants and $p_c$ is the critical occupation probability for classical (geometric) percolation. This result is to be contrasted with $P_{\infty}\propto(p-p_c)$ for $p\ge p_c$ just above the transition for $k=0,1$. In other words, the geometric percolation transition is a random first-order transition, i.e. a discontinuous transition with several diverging lengthscales.
\begin{figure}[bth]
\begin{center}
\includegraphics[width=6cm]{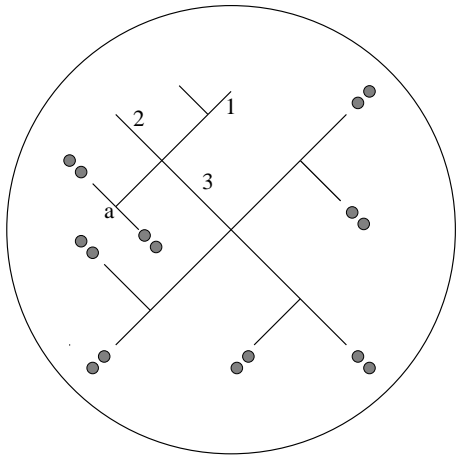}
\caption{Here $k=3$ and $z=4$.  The shaded circles denote branches
that are $k-1$ connected to infinity.  The removal of bonds 1 and
2 eventually triggers removal of bond 3 and bonds emanating from
vertex a, including the shaded circles. The remaining three
branches emanating from the center site survive the removal
process. } \label{fig:k3z4bethe}
\end{center}
\end{figure}

Once the $k$-core constraint has been implemented, we then impose the quantum percolation property that electrons can only hop along occupied bonds and compute the quantum conduction through the system self-consistently after assuming that the electronic wavefunction randomizes between bonds, i.e. the lengthscale of the disorder, $l_d$, is much greater than the coherence length, $l_c$.  Within this scheme,  for $k=0,1$, the model reduces to ordinary quantum percolation. In this case, we found that $1>p_q>p_c$,where $p_c$ signals the onset
of the geometric percolation transition and $p_q$ signals the onset of quantum conduction.  We also found a
random first-order transition for $k>2$, and that interestingly enough, the critical threshold is
the same as $k$-core geometrical percolation critical point. This transition should be contrasted with the Anderson-type MIT, which yields a continuous transition as does the quantum percolation transition on the Bethe lattice.

In this work, we numerically investigate the level statistics of
quantum $k$-core percolation to compare with our previous
analytical results on the Bethe lattice obtained in the limit
$l_d>>l_c$. In other words, how robust are our previous results in
identifying a new universality class of disorder-driven MITs?
Level statistics and its roots in random matrix theory is an
important tool for studying universality, for
example~\cite{mehta}. Correlations between energy eigenvalues of
an individual quantum particle in a random potential in the
conductive regime agrees with results from Gaussian matrix
ensembles~\cite{altshuler}. In the localized regime, correlations
between the energy eigenvalues are absent and the level statistics
become Poissonian. Right at the MIT, however, the level statistics
are distinct from Gaussian matrix ensembles~\cite{shklovskii}.
While these results pertain to the Anderson model where the
disorder is on-site, the same analysis has been applied to quantum
percolation on a cubic lattice, where the disorder is
off-diagonal, and similar results have been
found~\cite{berkovits}. In fact, the critical exponent for the
divergence of the localization length extracted from the level
statistics analysis is somewhat consistent with the Anderson
model. Therefore, we will implement a similar numerical analysis
for quantum $k$-core percolation on a Bethe-like lattice to go
beyond our previous approximation~\cite{liang.jen}.

We, however, will not investigate the level statistics of quantum $k$-core percolation on finite-dimensional lattices for now. It turns out that $k$-core percolation on finite-dimensional lattices either exhibits properties of $k=0$ geometric percolation or no transition. See, for example, Ref.~\cite{brazilians}. One has to invoke more sophisticated constraints to observe different universal behavior from $k=0$ geometric percolation~\cite{momo.jen} and so we expect the quantum behavior to be similar to quantum percolation for finite-dimensional lattices, though this conjecture should ultimately be tested.

The paper is organized as follows. We review the methodology and results for quantum percolation on the cubic lattice~\cite{berkovits} as a means for calibration and then present our results for quantum $k=0$-core percolation and quantum $k=3$-core percolation on a Bethe-like lattice. We conclude with a discussion of the implications of our results.

\section{Quantum percolation on the cubic lattice}

First, we analyze the level statistics for quantum percolation on the cubic lattice. To do so, we
diagonalize the Hamiltonian defined by Eqs. 1 and 2 on the cubic lattice of length $L$ with periodic boundary conditions to obtain a sequence of eigenvalues. This sequence is calculated for different realizations and increasing system sizes. The average density of states (DOS) for
$L=15$ as a function of occupation
probabilities is presented in Fig. 2. The sharp peaks are due to small disconnected structures as discussed in Ref.~\cite{berkovits}. Now, one can
apply the various measures of level statistics only if the density
of states is smooth. There exists a smooth energy range around 0.4.  The eigenenergies near this range are then
arranged from highest to lowest and the nearest-neighbor level spacing, $S$, is
calculated and subsequently normalized by the average nearest-neighbor level spacing, i.e. normalized level spacing is denoted by $s=S/<S>$.

One can then study the probability distribution for these level spacings, $P(s)$, as a function of $p$ and $L$. When the system is in the insulator regime, the eigenfunctions are
localized and, therefore, do not interact with each other such that $P(s)$ is Poisson distributed, i. e.
$P(s)=e^{-s}$.  When the system is in the metallic regime, Altshuler and
Shklovskii~\cite{altshuler} argued that if the width of the energy
band of a sample $E<E_c\equiv{\hbar}D/L^{2}$, where $L^{2}/D$
is the characteristic time for an electron to diffuse through the
sample, then the Hamiltonian of the system is characteristic of
Gaussian Orthogonal Ensemble (GOE) in the absence of a magnetic field or spin-orbit scattering. More specifically, $P(s)$ obeys the Wigner-Dyson
distribution, $P(s)=\frac{{\pi}s}{2}exp(-\frac{\pi}{4}s^2)$. The
cubic conductance $<G>=\frac{e^{2}}{\hbar}<N(E_c)>=\frac{e^{2}}{\hbar}\frac{E_c}{<S>}$
tends to infinity when $L\rightarrow\infty$ in the metallic
regime, thus $E<E_c$ is satisfied, and the
level spacing distribution obeys the Wigner-Dyson distribution for the GOE.

The plot of $P(s)$ as function of bond occupation probability for
$L=15$ and an energy range of $[0.2,0.6]$ is displayed in
Fig. 2. It can be seen that
the expected transition from Wigner-like behavior for large $p$
to Poisson behavior for small $p$. Note that all curves
intersect at $s\simeq2$ as observed in the Anderson model~\cite{shklovskii}. Other energy ranges studied yielded similar results. A convenient way to obtain the critical exponent $\nu$, characterizing the diverging localization length at the transition, is to study the parameter,
${\gamma}$, defined as \be
{\gamma}=\frac{\int_2^{\infty}P(s)ds-e^{-{\pi}}}{e^{-2}-e^{-{\pi}}},
\label{eq:gamma}\ee which characterizes the transition from
Wigner to Poisson as $p$ is decreased. Note that $\gamma$ should increase from 0 to 1, as $P(s)$ goes from Wigner to Poisson. Denoting ${\xi}(p)$ as the localization length (such that $\xi(p)\sim(p-p_q)^{-\nu}$),
the above parameter is expected to demonstrate scaling behavior,
${\gamma}(p,L)=f[L/{\xi}(p)]$. In the vicinity of the
critical quantum bond probability $p_q$,
\be {\gamma}(p,L)={\gamma}(p_q,L)+C|\frac{p}{p_q}-1|L^{1/{\nu}},
\label{eq:gamma.2}\ee
where $C$ is a constant.

Figure 3 plots $\gamma$ as a function of $p$ around $p_q$
for increasing $L$. The curves intersect near a single point given by
$p_q\cong0.334$. This result is consistent with the result in Ref.~\cite{berkovits}. The
single crossing point indicates that there exists a metal-insulator transition and one can apply the scaling collapse suggested above. The optimal scaling collapse shown in Fig. 3 yields $\nu=1.6\pm0.05$, which should be compared to $1.32\pm0.08$ obtained in Ref.~\cite{berkovits}.  Our latest result is even closer to the Anderson result than the previous work, where the most precise measurement is $\nu=1.58\pm 0.02$~\cite{andersonexp}.

\vspace{0.5cm}
\begin{figure}[bth]
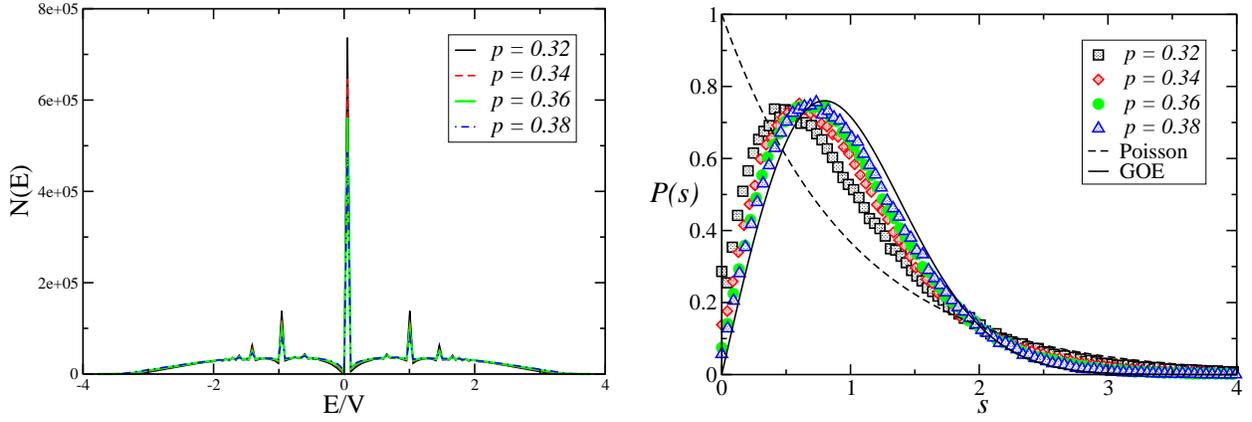

\begin{center}
\includegraphics[width=8cm]{dos.xm.eps}\label{fig:dos}
\hspace{0.2cm}
\includegraphics[width=8cm]{qp.level.spacing.dist.3d.xm.eps}\label{fig:3d}
\caption{Left: The DOS for $L=15$ and different bond occupation probabilities. Right: $P(s)$ for $L=15$ for different $p$s. The Wigner and Poisson forms are also shown.}
\end{center}
\end{figure}

\begin{figure}[bth]
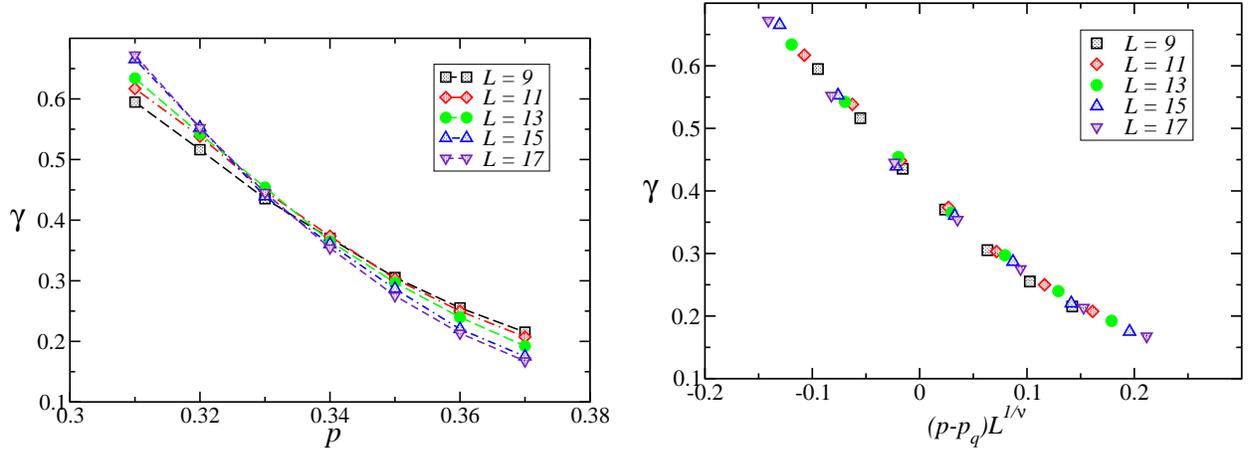

\begin{center}
\includegraphics[width=8cm]{qp.gamma.3d.xm.eps}
\hspace{0.2cm}
\includegraphics[width=8cm]{qp.gamma.scaling.3d.xm.eps}
\caption{ Left: The function $\gamma(p,L)$ for different system
sizes. Right: Scaling collapse for the cubic lattice.}
\end{center}
\end{figure}

\section{Quantum $k$-core percolation on a Bethe-like lattice}

Following the procedure presented in Ref.~\cite{bethelattice}, a Bethe-like lattice is numerically created by first considering a one-dimensional ring with $N$ sites such that each lattice site has two bonds emanating from it. Next, additional bonds are constructed between different lattice sites at random. The number
of random pairs connecting different lattice sites depends on the fixed
coordination number, $z$. More specifically, there must be $z$ bonds per site. See Fig. 4 as an
example for $z=3$.  As $N$ increases, the average number of loops of length $l$ increases as $(z-1)^l$ such that the fraction of all lattice sites belonging to any loop of length $\le l$ for $l<<\log(N)/\log(z-1)$ is negligible.  Therefore, the structure becomes increasingly tree-like as $N$ increases and no surface effects that have to be dealt with, which occurs when one performs numerical simulations on the Bethe lattice.

\subsection{$k=0$}

For $k=0$, A. B. Harris~\cite{harris} gave a theoretical
prediction of $p_q$ on the Bethe lattice. One must simply solve
$1+(p_q{\sigma}^2)^{-1}=(p_q{\sigma})^{2/({\sigma}-1)}$ with
${\sigma}=z-1$.  By solving Harris's self-consistency equation
above for $z=6$, $p_q=0.265$. We can test this result numerically
with the level statistics analysis. Figure 5 plots $P(s)$ for
different $p$s with $N=1024$, $z=6$ and an energy range of
$[0.3,0.8]$.  Figure 6 plots $\gamma(p,N)$ for $z=6$. The curves
intersect near $p\sim 0.3$. which is close to the analytical
result for the Bethe lattice, though the agreement is not precise.
This difference between the analytical calculation and our
numerical calculation is presumably due to the nature of the
lattice such that the quantum mechanics is much more sensitive to
loops than geometric percolation. To test this notion, when
measuring the onset of geometric percolation on the Bethe-like
lattice, we do arrive at good agreement between the analytical
result, $p_c=1/(z-1)$, and our numerical result.

Moreover, the
crossing point in Figure 6 indicates that we can collapse the data by
assuming that, instead of ${\gamma}(p,L)=f[L/{\xi}(p)]$ in the
cubic lattice case, ${\gamma}(p,L)=f[N/N^*(p)]$, where
$N^*$ is a crossover size similar to the localization length in the
three-dimensional case. In other words, $N^*\sim(p-p_q)^{-\nu'}$. Therefore, ${\gamma}(p,L)=f[N^{1/{\nu}'}(p-p_q)]$. It has been conjectured that $\nu'=d_u\nu_{MF}$, where $d_u$ is the upper critical dimension and $\nu_{MF}$ is the mean field correlation length exponent~\cite{networks}.  For geometric percolation, $d_u=6$ and $\nu_{MF}=1/2$ such that $\nu'=3$.  To date, the upper critical dimension of quantum percolation is not known.  Interestingly, the upper critical dimension for the Anderson model is potentially infinite such that the mean field correlation length scales with dimension $d$~\cite{infinite}. More precisely, $\nu=\frac{1}{2}+\frac{1}{d-2}$.

Figure 6 shows the optimal scaling collapse to yield the exponent
${\nu}'=4.5\pm0.2$ assuming that $p_q=0.300(1)$.  If we assume
$p_q=0.265$, then we do not arrive at a good scaling collapse.
Regarding $\nu'$, if we assume the same mean field correlation
length of $\nu_{MF}=1/2$ as in classical $k=0$ percolation, which
is consistent with the previous work of A. B. Harris~\cite{harris}
showing that the mean field susceptibility for zero-energy
eigenstates diverges with the same exponent as in the geometric
percolation problem (with $p_q$ replacing $p_c$), then we extract
an upper critical dimension of $d_u=9$.  If, on the other hand,
the upper critical dimension is infinite as in the case of the
Anderson model, then different analysis must be undertaken.

\begin{figure}[bth]
\begin{center}
\includegraphics[width=8cm]{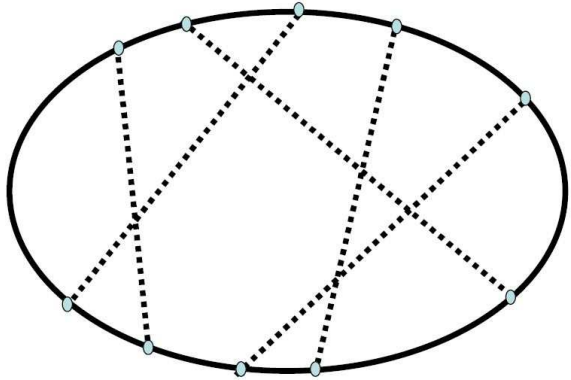}
\caption{An example of a Bethe-like lattice with coordination
number, $z=3$. Dotted curves indicate the random pairs.}
\label{fig:bethelattice.fig}
\end{center}
\end{figure}
\vspace{1cm}
\begin{figure}[bth]
\begin{center}
\includegraphics[width=8cm]{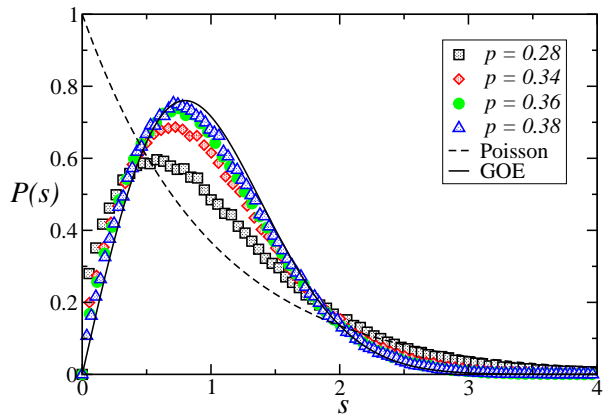}
\caption{ $P(s)$ for the Bethe-like lattice for
$N=1024$ and $k=0$ for different $p$s.}
\end{center}
\end{figure}

\begin{figure}[t]
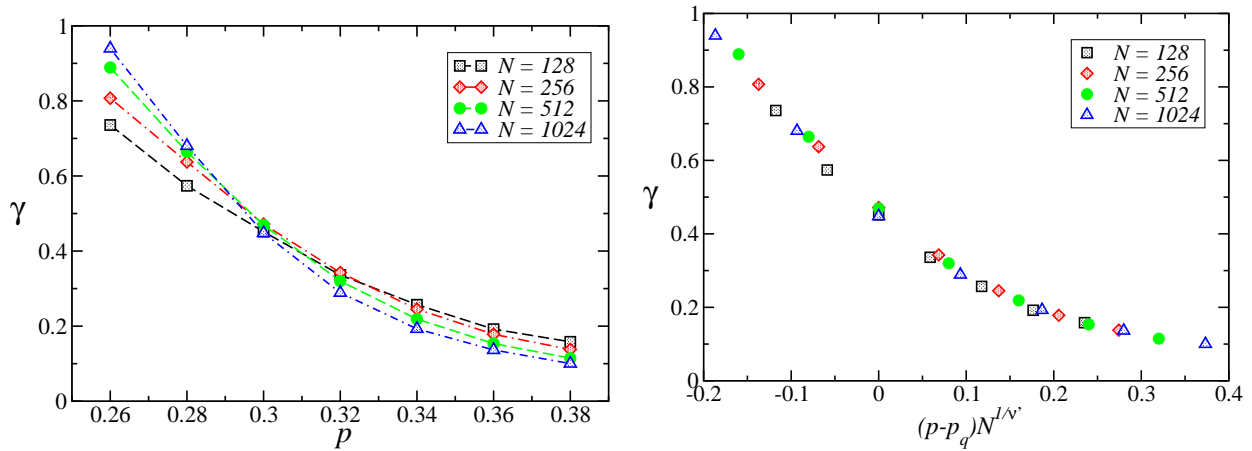

\begin{center}
\includegraphics[width=8cm]{qp.gamma.k0.bethe.xm.eps}
\hspace{0.2cm}
\includegraphics[width=8cm]{qp.gamma.scaling.k0.bethe.xm.eps}
\caption{Left: The function ${\gamma}(p)$ for different system sizes. Right: The scalling collapse for ${\gamma}(p)$.}
\end{center}
\end{figure}

\subsection{$k=3$}

In Ref.~\cite{liang.jen}, we gave an example of a MIT driven by $k=3$-core disorder on the Bethe lattice with $z=4$. More precisely, the quantum conduction as a function of
occupation probability $p$ is a random first-order transition with
$p_q=p_c$. To obtain this result, we assumed that the phase randomized between levels on the Bethe lattice, i.e. $l_d>>l_c$.  Here, we use level
statistics to test the robustness of our prior results. Since the geometric critical
percolation occupation probability $p_c$ is 8/9, which is close to
1, we choose $z=6,k=3$, whose $p_c=0.603$~\cite{clr,network,slc}. Numerically,
Figure 7 gives the fraction of
remaining samples versus occupation probability, denoted by $F$, after applying the
$k$-core constraint recursively for $z=6, k=3$. One can find
that in a ideally infinite system size system, all occupied bonds are
removed due to the $k$-core constraint for $p<~0.604$, while $p>~0.604$, occupied bonds remain.  Note that $p_c=p_q=0.604$ is quite
close to the analytical $p_c$ for $z=6$ and
$k=3$ for the classical physics.

Figure 8 plots ${\gamma}$ as a function of
$p$ for different system sizes. As the system size increases, larger
error bars are shown for $p<p_c$, indicating less remaining
occupied samples due to the $k$-core constraint. With no remaining occupied bonds for $p<p_c$, the system is insulating in a trivial sense and not in the sense that $\gamma\rightarrow 1$. What happens for $p>p_c$?  The data indicates that ${\gamma}$ tends to 0 as the system size increases for $p>p_c$, indicating that it is a metallic system. There is no crossing point above $p_c$ such that $p_c$ must equal $p_q$ as we obtained previously.  Note that we could have observed a crossing point somewhere above $p_c$ to indicate that $p_q>p_c$, but we do not. The data imply a discontinuous transition since $\gamma$ tends towards zero as $N$ becomes large for $p\ge p_q$.

\begin{figure}[b]
\begin{center}\vspace{0.5cm}
\includegraphics[width=8cm]{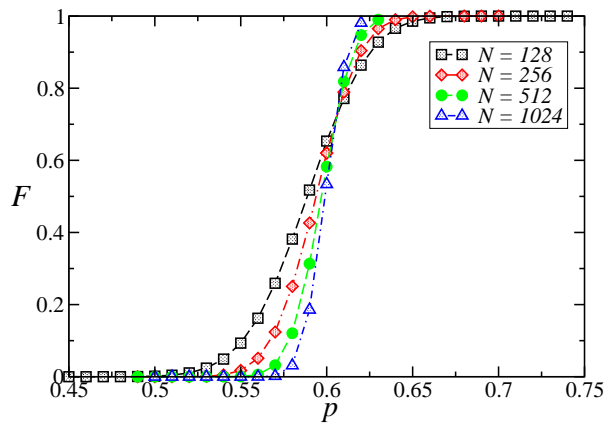}
\caption{Fraction remaining samples versus occupation
probability after applying the $k$-core constraint, $F$, on the Bethe-like lattice for
$z=6,\,k=3$.} \label{k-coreremainingsample}
\end{center}
\end{figure}

\section{Discussion}
We have applied level statistics for quantum percolation on three
different systems. The results on cubic lattice are consistent
with the results from Ref.~\cite{berkovits}. For quantum
percolation on a Bethe-like lattice, we find a threshold
probability that differs somewhat from Harris's Bethe lattice
result~\cite{harris}. This difference is due to the sensitivity of
loops in quantum mechanics occurring in the Bethe-like lattice.
Moreover, a scaling collapse with system size $N$ yields a new
critical exponent $\nu^{\prime}=4.5(2)$. Assuming the correlation
length exponent is $\nu_{MF}$ is 1/2, the same as the mean field
result for geometric percolation, which is consistent with the
Harris calculation~\cite{harris}, the upper critical dimension
$d_u=9$. We have also demonstrated the robustness of our previous
work for quantum $k=3$-core percolation~\cite{liang.jen}. The
level statistics here shows $p_q=p_c$, as before. In fact, for
$p<p_c$, all bonds are removed and the system is an insulator,
while for $p>p_c$, none of the bonds are removed and the system is
conducting quantum mechanically. The data suggest that the system
immediately goes to the Wigner-Dyson regime without going through
a different regime at the transition, indicating a discontinuous
transition, which also agrees with our previous
work~\cite{liang.jen}.

\begin{figure}[bth]
\begin{center}
\includegraphics[width=8cm]{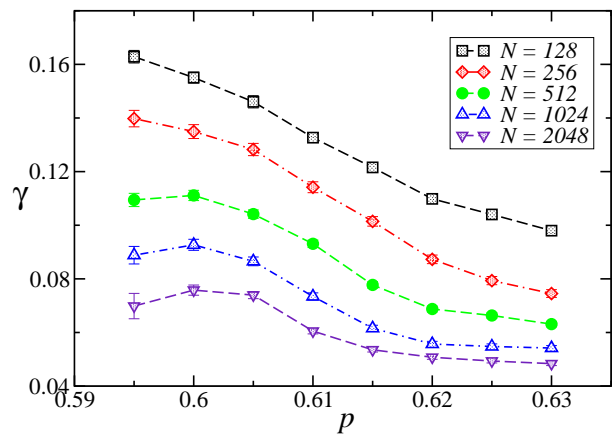}
\caption{${\gamma}$ as a function of $p$ for different system
sizes for the $z=6,k=3$ Bethe-like lattice.} \label{gamma.k=3}
\end{center}
\end{figure}

Therefore, our work provides an important counterexample for the Mott versus Anderson MIT paradigm, where disorder-driven (Anderson) MITs are continuous and
interaction-driven (Mott-Hubbard) MITs are discontinuous. Our counterexample is due to the correlations in the disorder as a result of the $k$-core constraint. Correlations in the disorder have been shown previously to ``complicate'' matters.  For instance, in one-dimensional wires with long-range, correlated disorder, there exists a MIT, which does not happen in the short-range, uncorrelated disorder case~\cite{oned}.

As for experimental implications, there are many experiments in the realm of classical transport on ordinary percolating systems.  See, for example, Ref.~\cite{nanowire1} for a recent one on nanowire composites. In addition, transport in undoped graphene is linked to classical electronic transport on percolating networks, though quantum effects are also relevant~\cite{graphene}. However, can a quantum $k=3$-core percolation transition ever be realized?  An experiment has
already been conducted with
a collection of silver quantum dots sitting atop of a Langmuir monolayer at
room temperature~\cite{heath}.  As the
interparticle spacing decreases by compressing the floating particles together, the electronic
transport goes from hopping to tunnelling to ordinary
metallic transport.
The authors argue that disorder in the particle size and in the
charging energy
probably does not drive the transition and, instead, conjecture a possible
first-order Mott transition at {\em room temperature}. However, in light of the analysis of the onset of classical conduction for $k=3$-core, where $k$ encodes the scalar aspect of local mechanical stability in particle packings~\cite{slc}, we
argue for a possible {\em classical} correlated percolation transition
in conduction. Perhaps a quantum analog of this experiment can be realized in low-temperature packings of metallic nanoparticles to search for this new universality class of quantum disorder-driven MITs.

JMS kindly acknowledges funding support from NSF-DMR-CAREER-0645373.


\begin{thebibliography}{10}

\bibitem{mott} N. F. Mott, ``Metal-insulator transition,'' {\it Rev. Mod. Phys.} {\bf 40}, 677 (1968).

\bibitem{anderson}  P. W. Anderson, ``Absence of diffusion in certain random lattices,'' {\it Phys. Rev.} {\bf 109}, 1942 (1958).

\bibitem{mirlin} F. Evers and A. D. Mirlin, ``Anderson transitions,'' {\it Rev. Mod. Phys.} {\bf 80}, 1355 (2008).

\bibitem{liang.jen} L. Cao and J. M. Schwarz, ``Quantum $k$-core conduction on the Bethe lattice ,'' {\it Phys. Rev. B} {\bf 82}, 104211
(2010).

\bibitem{degennes} P. G. de Gennes, P. Lafore and J. P. Millot, ``Amas accidentels dans les solutions solides d\'esordonn\'ees,'' {\it J. Phys. Chem. Solids} {\bf 11}, 105 (1959).

\bibitem{germans2} G. Schubert and H. Fehske, ``Quantum percolation in
  disordered structures,'' {\it Lect. Notes Phys.} {\bf 762}, 135 (2009).

\bibitem{germans1} G. Schubert, A. Weibe, H. Fehske, ``Localization effects in quantum percolation,'' {\it Phys. Rev. B} {\bf 71}, 045126 (2005).

\bibitem{travenec} I. Travenec, ``Metal-insulator transition in 3d quantum percolation,'' {\it Int. J. Mod. Phys. B} {\bf 22}, 5217 (2008).

\bibitem{kaneko} A. Kaneko and T. Ohtsuki, ``Three-dimensional quantum percolation studied by level statistics,''  {\it J. Phys. Soc. Jpn.} {\bf 68}, 1488 (1999).

\bibitem{meir} Y. Meir, A. Aharony, and A. B. Harris, ``Delocalization transition in two-dimensional quantum percolation,'' {\it Europhys. Lett.} {\bf 10}, 275 (1989).


\bibitem{chang} D. Daboul, I. Chang, and A. Aharony, ``Series expansion study of quantum percolation on the square lattice,'' {\it Eur. Phys. J. B} {\bf 16} 303 (2000).

\bibitem{islam1} M. F. Islam and H. Nakanishi, ``Localization-delocalization transition in a two-dimensional quantum percolation model,'' {\it Phys. Rev. E} {\bf 77}, 061109 (2008).

\bibitem{islam2} H. Nakanishi and M. F. Islam, ``Quantum percolation in two dimensions,'' {\it Lect. Notes Phys.} {\bf 762}, 109 (2009).

\bibitem{germans3} G. Schubert and H. Fehske, ``Dynamical aspects of two-dimensional quantum percolaton'', {\it Phys. Rev. B} {\bf 77}, 245130 (2008).

\bibitem{grest} C. M. Soukoulis and G. S. Grest, ``Localization in two-dimensional quantum percolation'', {\it Phys. Rev. B} {\bf 44}, 4685 (1991).

\bibitem{mookerjee} A. Mookerjee, I. Dasgupta, and T. Saha, ``Quantum percolation'', {\it Int. J. Mod. Phys. B} {\bf 9}, 2989 (1995).

\bibitem{harris0} A. B. Harris, ``Exact solution of a model of localization'', {\it Phys. Rev. Lett.} {\bf 49}, 296 (1984).

\bibitem{harris} A. B. Harris, ``$1/{\sigma}$ expansion for quantum percolation,'' {\it Phys. Rev. B} {\bf 29}, 2519
(1984).


\bibitem{clr}  J. Chalupa, P. L. Leath and G. R. Reich, ``Bootstrap percolation on a Bethe lattice,''   {\it J. Phys. C: Solid St. Phys.} {\bf 12},
L31 (1979).

\bibitem{wormald} B. Pittel, J. Spencer, and N. Wormald, ``Sudden emergence of a giant $k$-core in a random graph,'' {\it J. Comb. Th. Ser. B} {\bf 67}, 111 (1996).

\bibitem{network0} S. N. Dorogovtsev, A. V. Goltsev, and J. F. F. Mendes, ``$k$-core organization of complex networks'', {\it Phys. Rev. Lett.} {\bf 96}, 040601 (2006).

\bibitem{network} A. V. Goltsev, S. N. Dorogovtsev, and J. F. F. Mendes, ``$k$-core (bootstrap) percolation on complex networks: Critical phenomena and nonlocal effects,'' {\it Phys. Rev. E} {\bf 73}, 056101 (2006).

\bibitem{slc} J. M. Schwarz, A. J. Liu, and L. Q. Chayes, ``The onset of jamming as the sudden emergence of an infinite $k$-core cluster,'' {\it Europhys. Lett.} {\bf 73}, 560 (2006).

\bibitem{harris.schwarz} A. B. Harris and J. M. Schwarz, ``$1/d$ expansion in $k$-core percolation,'' {\it Phys. Rev. E} {\bf 72}, 046123 (2005).

\bibitem{biroli} M. Selitto, C. Toninelli, and G. Biroli, ``Facilitated spin models on Bethe lattice: Bootstrap percolation, mode coupling transition and glassy dynamics'', {\it Europhys. Lett.} {\bf 69}, 496 (2005).

\bibitem{levine} D. J. Schwab, R. F. Bruinsma, J. L. Feldman and A. J. Levine, ``Rhythmogenic neuronal networks, emergent leaders, and $k$-cores'', {\it Phys. Rev. E} {\bf 82}, 051911 (2010).

\bibitem{mehta} M. L. Mehta, {\it Random Matrices} (Academic Press, San Diego, 1991), and the references therein.


\bibitem{altshuler} B. L. Altshuler and B. I. Shkolvskii, ``Repulsion of energy levels and conductivity of small metal
samples,'' {\it Zh. Eksp. Teor. Fiz.} {\bf 91}, 220 (1986)[Sov. Phys.
JETP {\bf 64},127 (1986)].

\bibitem{shklovskii} B. I. Shklovskii, B. Shapiro, B. R. Sears, P. Lambrianides, and H. B. Shore, ``Statistics of spectra of disordered systems near the metal-insulator
transition,'' {\it Phys. Rev. B} {\bf 47}, 11487 (1993).

\bibitem{berkovits} R. Berkovits and Y. Avishai, ``Spectral statistics near the quantum percolation
threshold,'' {\it Phys. Rev. B} {\bf 53}, R16125 (1996).

\bibitem{brazilians} M. C. Medeiros and C. M. Chaves, Physica A {\bf
234},
604 (1997); C. M. Chaves and B. Koiller, {\it Physica A} {\bf 218}, 271
(1995).

\bibitem{momo.jen} M. Jeng and J. M. Schwarz, {\it Phys. Rev. E} {\bf 81}, 011134 (2010).

\bibitem{andersonexp} K. Slevin and T. Ohtsuki, ``Corrections to Scaling at the Anderson transition'', {\it Phys. Rev. Lett.} {\bf 82}, 382 (1999).

\bibitem{bethelattice}  D. Dhar, P. Shukla, and J. P. Sethna, ``Zero-temperature hysteresis in the random-field Ising model on a Bethe lattice,'' {\it J. Phys. A: Math. Gen.} {\bf 30}, 5259 (1997).

\bibitem{networks} Y. Kim, Y Ko, and S.H. Yook, ``Structural properties of the synchronized cluster on complex networks'', {\it Phys. Rev. E} {\bf 81}, 0111139 (2010).

\bibitem{infinite} A. M. Garcia-Garcia, ``Semiclassical theory of the Anderson transition'', {\it Phys. Rev. Lett.} {\bf 100}, 076404 (2008).

\bibitem{oned} P. Carpena, P. Bernaola, P. C. Ivanov, and H. E. Stanley, ``Metal-insulator transition in chains with correlated disorder'', {\it Nature} {\bf 418}, 955 (2002).


\bibitem{nanowire1} S. I. White, R. M. Mutiso, P. M. Vora, D. Jahnke, S. Hsu,
J. M. Kikkawa, J. Li, J. E. Fischer, and K. I. Winey,
``Electrical Percolation Behavior in Silver Nanowire¨CPolystyrene
Composites: Simulation and Experiment'', {\it Adv. Funct. Mater.} {\bf
20}, 2709 (2010).

\bibitem{graphene} V. V. Cheianov, V. I. Fal'ko, B. L. Altshuler, and I. L. Aleiner, ``Random resistor network model of minimal conductivity in graphene'', {\it Phys. Rev. Lett.} {\bf 99}, 176801 (2007).

\bibitem{heath} G. Markovich, C. P. Collier, and J. R. Heath, ``Reversible metal-insulator transition in ordered metal nanocrystal monolayers observed by impedance spectroscopy,'' {\it Phys. Rev. Lett.} {\bf 80}, 3807 (1998).
\end{thebibliography}
\end{document}